\documentclass[a4paper]{article}
\usepackage{graphicx}
\usepackage{color}
\begin{document}
\title{Q/V enhancement of Si micropillar resonator with Bragg reflectors in BIC regime}

\author{Stanislav Kolodny and Ivan Iorsh}
\maketitle


\begin{abstract}
We show how the recently found quasi-bound state in the continuum optical states in dielectric antennae can be utilized to engineer pillar microcavities design with the substantially optimized quality factor to mode volume ratio.
\end{abstract}

\section{Introduction}
Boosting light-matter interaction is of crucial importance for many fields of nanophotonics such as cavity quantum electrodynamics~\cite{Verger2006,Jarlov2013,Raftery2014,Hartmann2016}, optomechanics~\cite{RevModPhys.86.1391, PhysRevA.82.031804, Safavi2010}, and low-threshold lasers~\cite{Painter1999,Ota2013,Xue2016,Ellis2011,Wu2015}.
A natural figure of merit for the light-matter interactions is $F=\lambda^3Q/V$, where $Q$ is  cavity quality factor, $V$ - cavity mode volume and $\lambda$ is the resonant wavelength. Conventionally, large values of $F$ can be achieved in photonic crystal cavities~\cite{Wang2018}, open cavities~\cite{Dufferwiel2014} and pillar microcavities~\cite{Moreau2001}. The latter comprises a dielectric resonator sandwiched between the two Bragg mirrors. While the decay through the mirrors can be almost totally suppressed by increasing the number of periods, the main route of the radiation loss is via the side-walls. These can be decreased by exploiting the modes with large angular momentum, whispering gallery modes of the pillar microcavity~\cite{Nowicki-Bringuier:07}, which however will inevitably increase the effective mode volume. Recently, a novel route has been proposed to combine high quality factors and low mode volumes in dielectric resonators, based on the concept of bound state in the continuum (BIC)~\cite{rybin2017high}. Optical bound states in the continuum usually occur in periodic structures: these are the virtually lossless optical states occurring due to the destructive interference of the leaky modes at particular frequency and wavevector~\cite{PhysRevLett.100.183902}. In~\cite{rybin2017high} it has been shown, that similar phenomena can occur in compact systems, such as isolated dielectric resonators: in this case the destructive interference occurs between different Mie resonant modes; the linewidth of these quasi-BIC states remains finite, however at the resonant frequency the system effectively radiates at the higher angular momentum Mie resonance, thus simultaneously increasing the quality factor and almost unaffecting the effective mode volume. 

Here we employ this approach for the case of pillar microcavities. Namely, we show that just by tuning the height to radius ratio of the cavity we can achieve the destructive interference between the low energy fundamental resonant modes and substantially increase the $Q/V$ ratio.

\section{Result and Discussion}
\begin{figure}[h!]
\begin{center}
\includegraphics[width=0.95\linewidth]{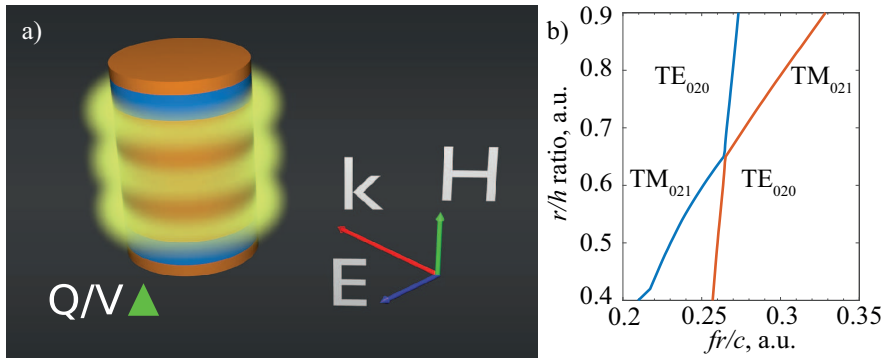}
\end{center}
\caption{\label{1st}a) Schematic presentation of Si micropillar resonator sandwiched by Bragg reflectors. Notice that for simplicity there are only 2 periods of Bragg mirrors presented. b) Anti-crossing of considered $TE_{020}$ and $TM_{021}$ modes of Si micropillar resonator. Y-axe represents radius/height aspect ratio of resonator and X-axe represents the normalized frequency $x = fr/c$. }
\end{figure}
\begin{figure}[h!]
\begin{center}
\includegraphics[width=\linewidth]{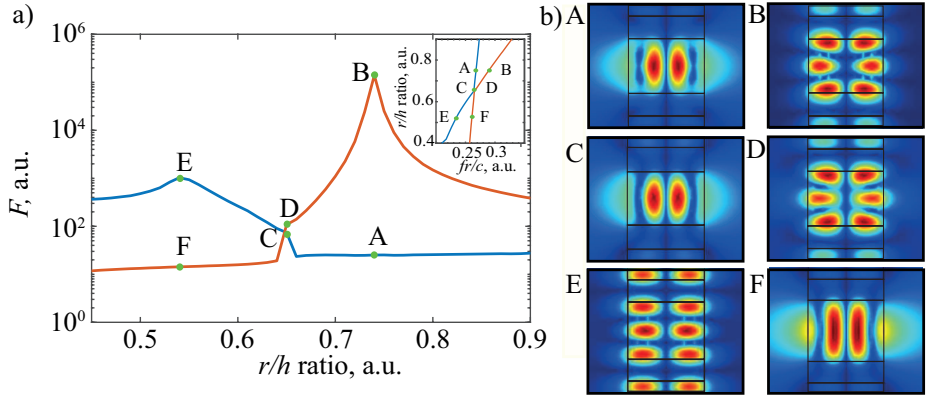}
\end{center}
\caption{\label{F_w}a) The Purcell factor of $TM_{021}$ and $TE_{020}$ modes of Si micropillar resonator sandwiched between 8 periods of DBM top and bottom. Blue line corresponds to left branch and orange corresponds to right branch which are shown on the inset as blue and orange line, respectively. b) Electric field distribution $|E|$ for the corresponding points indicated on the inset to Fig.\ref{F_w}~a).}
\end{figure}

In order to increase the Q/V ratio (also can be defined as well-known Purcell factor) we  use the distributed Bragg reflectors (DBR) placed on the top and bottom surfaces of Si micropillar resonator (see Fig.~\ref{1st}~a). Here we consider the hybridization of $TE_{020}$ and $TM_{021}$ modes with the same azimuthal number $m = 0$ (see Fig.~\ref{1st}~b). We perform  numerical simulations solving eigenmodes problem with varying $r/h$ aspect ratio of resonator while the radius of cylinder is fixed. The Purcell factor of an  eigenmode can be defined as:
\begin{equation}
    F = \left(\frac{\lambda}{n}\right)^3 \frac{Q}{V},
\end{equation}
where Q is a quality factor of the mode, V is an effective mode volume, $\lambda$ is the wavelength. In our simulations we look for the complex eigenfrequencies  $\omega = \omega'-i\omega''$, so
the Q-factor can be found from complex value of eigenfrequency as follows:
\begin{equation}
    Q = -\frac{Re(\omega)}{2Im(\omega)}
\end{equation}
The effective mode volume can be expressed as:
\begin{equation}
    V = \frac{\int_V \epsilon |E|^2 dV}{max(\epsilon |E|^2)},\label{modV}
\end{equation}
where integration is performed over the volume of the cavity and  all DBR periods in case of the pillar microcavity.
Thus, we can find the dependence of Purcell factor of the considered modes on $r/h$ aspect ratio and results are shown in the Fig.\ref{F_w}~a. Note that here and after the plots are given for left and right branches of anti-crossing regime as highlighted in color on the Fig.~\ref{1st}a and inset of Fig.~\ref{F_w}a. The structure
consists of a cylindrical silicon cavity and a top and bottom Bragg mirror, in which one of the layers is silicon with $\varepsilon_{Si}\approx 16$, and the other is silicon oxide with $\varepsilon\approx 2.1$. The DBR layers thicknesses are tuned such that the cavity frequency coincides with the DBR stop band center. Therefore, since the cavity frequency changes with the $r/h$ aspect ratio, the DBR layers thicknesses are changed accordingly. These changes are made independently for the two branches. The number of DBR periods has been chosen to be equal to 8 assuming that this will be enough to ensure almost complete suppression of radiation in the vertical direction. Naturally, in the case of the lower contrast DBRs, the number of the periods should be increased.
In  Fig.~\ref{F_w}~a we show the dependence of the  Purcell factor on the aspect ratio in the pillar microcavity with 8 DBR periods. In the frequency range considered, there are two resonant modes, supported by the cavity, TM$_{021}$ and TE$_{020}$, where the first index stands for the angular momentum of the mode, the second one for the radial wavenumber, and the third one for the wavenumber in the $z$ direction. In the pillar cavity with perfectly non-transmitting boundaries, these two modes would be orthogonal. However, due to the open nature of the cavity, these two modes become hybridized via the coupling in the outer region of the cavity. It can be noted from the $F$ spectra and field profiles in Fig.~\ref{F_w}, that TE$_{020}$ is defined by the low quality factors, mainly due to the leakage through the cavity side walls. Meanwhile, TM$_{021}$ mode primarily leaks through the Bragg mirror, almost does not penetrate through the side walls and is thus defined by relatively high Q-factors. Due to the leaky nature of the both modes and relatively low overlap of the electric field, the coupling between the modes does not lead to the avoided crossing behaviour of the modes' eigenfrequencies as can be seen in Fig.~\ref{1st}(b). However, a strong modification of the quality factors for the TM$_{021}$ is observed. Namely, at $r/h=0.741$ a strong resonant enhancement of the Purcell factor is observed by almost two orders of magnitude.   
\begin{figure}[h!]
\begin{center}
\includegraphics[width=\linewidth]{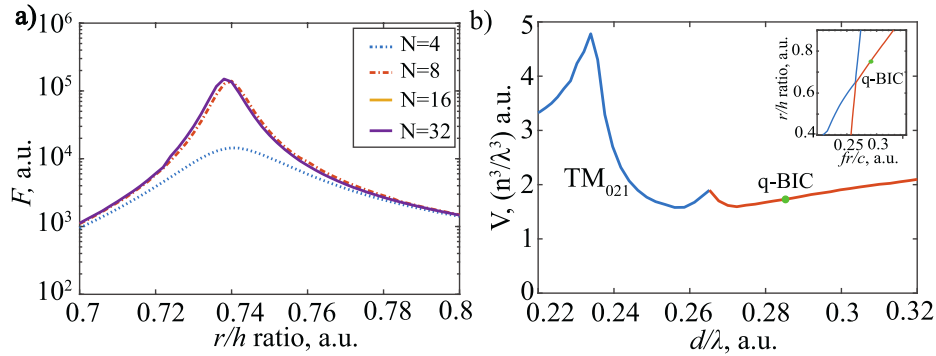}
\end{center}
\caption{\label{N}a) The Purcell factor of Si micropillar resonator in dependence on number of DBR periods (N) shown by different colors on the inset. The values of Purcell factor are given in decimal logarithmic scale. b) Normalized effective mode volume versus normalized diameter of resonator. The branches of anti-crossing are represented in color as it is presented in inset. The green point represents the position of BIC of $TE_{021}$ mode. }
\end{figure}
Increasing the number of DBR periods will increase the peak Purcell factor but up to some saturation value as can be seen in Fig.~\ref{N}(a). At saturation, the total decay of the mode is almost fully determined by the leakage through the side walls. For the considered case of the high contrast DBR structure, the saturation is established already at 8 DBR periods. For the lower contrast case (e.g. GaAs/AlGaAs DBRs) larger number of periods will be required. 

We note, that the overall effect of the resonant Purcell enhancement is achieved because at the q-BIC point, the Q-factor increases resonantly and the mode volume remains almost unchanged. In Fig.~\ref{N}(b) we plot the spectra of the resonant mode volume for the high quality mode. For the low quality mode we omit the spectra, since in this case the mode volume can not be properly evaluated by the simple Eq.~\ref{modV}. We can see that at the q-BIC point, the mode volume remains almost unchanched, which together with the sharp increase in quality factor facilitates the Purcell factor increase.

\section{Conclusion}
We have proposed a relatively simple method of the resonant enhancement of the Purcell factor in pillar microcavities by employing quasi-BIC compact modes. This method requires tuning of a single parameter, cavity radius to height ratio, does not require any additional  sophisticated fabrication techniques and allows to boost the Purcell factor by two orders of magnitude. We thus anticipate, that it can employed in the engineering of optoelectronic and quantum optical devices based on pillar microcavities.
\section*{Acknowledgments}
We acknowledge the support from the Ministry of Education and Science of Russian Federation (No. 3.1365.2017/4.6, No. 3.1500.2017/4.6 and No. 14.Y26.31.0015), Grant of President of Russian Federation MK-6248.2018.2 and RFBR Project 16-32-60123.

\providecommand{\newblock}{}

\end{document}